\documentclass[aps,prb,twocolumn,showpacs,superscriptaddress,floatfix]{revtex4-2}

\usepackage{graphicx}
\usepackage{amsmath}
\usepackage{dcolumn}
\usepackage{bm}
\usepackage{siunitx}
\usepackage{hyperref}
\hypersetup{colorlinks=true,citecolor={blue},linkcolor={blue},urlcolor={blue}}

\usepackage[caption=false]{subfig}
\newcommand{\phantomsubfloat}[1]{
    {
        \captionsetup[subfigure]{labelformat=empty}
        \subfloat[][]{#1}
    }
}
\usepackage{xcolor}
\usepackage{float}

\begin{document}

\title{Reservoir optics with exciton-polariton condensates}

\author{Y. Wang}
\address{School of Physics and Astronomy, University of Southampton,  Southampton, SO17 1BJ, UK.}

\author{H. Sigurdsson}
\email[H. Sigurdsson ]{h.sigurdsson@soton.ac.uk}
\affiliation{School of Physics and Astronomy, University of Southampton,  Southampton, SO17 1BJ, UK.}
\affiliation{Science Institute, University of Iceland, Dunhagi 3, IS-107, Reykjavik, Iceland}

\author{J. D. Töpfer}
\address{School of Physics and Astronomy, University of Southampton,  Southampton, SO17 1BJ, UK.}
\address{Skolkovo Institute of Science and Technology, Moscow, Territory of innovation center “Skolkovo”,
Bolshoy Boulevard 30, bld. 1, 121205, Russia.}

\author{P. G. Lagoudakis}
\address{Skolkovo Institute of Science and Technology, Moscow, Territory of innovation center “Skolkovo”,
Bolshoy Boulevard 30, bld. 1, 121205, Russia.}
\address{School of Physics and Astronomy, University of Southampton,  Southampton, SO17 1BJ, UK.}

\begin{abstract}
We investigate an all-optical microscale planar lensing technique based on coherent fluids of semiconductor cavity exciton-polariton condensates. Our theoretical analysis underpins the potential in using state-of-the-art spatial light modulation of nonresonant excitation beams to guide and focus polariton condensates away from their pumping region. The nonresonant excitation profile generates an excitonic reservoir that blueshifts the polariton mode and provides gain, which can be spatially tailored into lens shapes at the microscale to refract condensate waves. We propose several different avenues in controlling the condensate fluid, and demonstrate formation of highly enhanced and localised condensates away from the pumped reservoirs. This opens new perspectives in guiding quantum fluids of light and generating polariton condensates that are shielded from detrimental reservoir dephasing effects.
\end{abstract}

\maketitle

\section{Introduction}

Advancements in guiding and focusing the flow of planar (paraxial) light waves at the microscale brings far-reaching possibilities into miniaturized optical technologies, from microlens arrays~\cite{Yuan_ChinJourME2018} to optical circuitry and logic gates~\cite{Singh_Hindawi2014}, that are reliant on dispersion management. Metamaterials~\cite{lu2012hyperlenses, Khorasaninejad_Science2017}, plasmonic lenses~\cite{Liu_NanoLett2005, Kim_OptExp2008, Verslegers_NanoLett2009}, phase-change materials~\cite{chen2015engineering}, photonic crystals~\cite{Parimi_Nature2003, Casse_APL2008} and disordered materials~\cite{Leonetti_NatComm2014} all offer a variety of techniques to focus planar light, though usually coming at the cost of irreversible fabrication methods. Here, we introduce an all-optical planar microlensing approach in a system of microcavity exciton-polariton condensates that offers flexible and reprogrammable lens configurations.

\begin{figure}[t]
\includegraphics[scale=1.0]{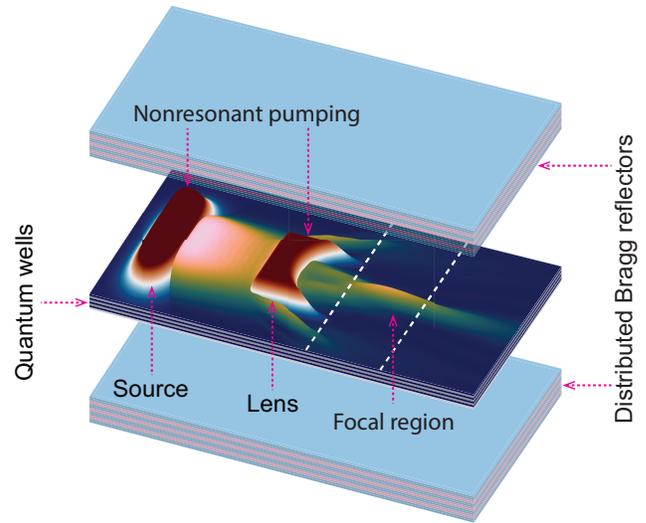}
\caption{Sketch of the nonresonant lensing effect with exciton-polariton condensates in a semiconductor microcavity. Quantum well excitons are photoexcited through nonresonant pumping (dark red profiles) and while the cavity mirrors (symmetric distributed Bragg reflectors) provide photon confinement and strong-coupling.}
\label{fig1}
\end{figure}

Exciton-polaritons (from here on {\it polaritons}) are boson-like quasiparticles formed by coherent hybridization of electron-hole pairs in semiconductor quantum wells and microcavity photons in the strong-coupling regime~\cite{RevModPhys.85.299}, as sketched in Fig.~\ref{fig1}. The extremely small effective polariton mass ($\sim 10^{-5}$ of the electron mass) and large interaction strength, due to the excitonic component, has opened up new strategies in all-optical control over macroscopic coherent matter-wave fluids of light, or polariton condensates~\cite{kasprzak2006bose}. For the past ten years there have been several important experiments in all-optical manipulation of polariton condensates using nonresonant excitation methods~\cite{Wertz_NatPhys2010, Tosi_NatPhys2012} such as condensate amplification~\cite{PhysRevLett.109.216404}, trapping~\cite{Cristofolini_PRL2013, Askitopoulos_PRB2013}, exceptional points~\cite{Gao_Nature2015}, dissipative annealing of the XY model~\cite{Berloff_NatMat2017}, vortex manipulation~\cite{PhysRevLett.113.200404, Ma_NatComm2020}, and lattices~\cite{Pickup_NatComm2020, Pieczarka_Optica2021}. Many works have also combined the optical control provided by nonresonant lasers in conjunction with engineered photonic potentials such as micropillars, microwires, or wedged cavities (i.e., photonic potential gradient) which led to development of optically controllable interferometers~\cite{Sturm_NatComm2014} and transistor switches~\cite{Gao_PRB2012, Zasedatelev_NatPho2019}. Alongside these developments in optical control, there is a growing variety in cheaper room-temperature materials that operate in the strong coupling regime~\cite{Christopoulos_PRL2007, sanvitto2016road, Su_SciAdv2018} which opens new perspectives on the role of exciton-polaritons in future optical based technologies~\cite{Zasedatelev_NatPho2019}.

In this study, we explore spatial control over propagating exciton-polariton condensates using nonresonant excitation beams shaped into a planoconcave microlens (see Fig.~\ref{fig1}). The excitation beam induces (photoexcites) a static reservoir of incoherent excitons which provide both gain and blueshift to the polariton modes~\cite{Wertz_NatPhys2010, Tosi_NatPhys2012}. Consequently, excited polaritons experience a complex valued effective potential landscape which can both amplify and phase-modulate transmitted waves. When the excitation beam is removed the reservoir rapidly decays which permits rewriting new and different potential landscapes on the same sample location. Recently, similar flexibility was demonstrated with phonon-polaritons in hexagonal boron nitride heterostructure~\cite{Chaudhary_NatComm2019}. So far, there have been several studies addressing the potential in nonresonant all-optical control to manipulate the flow of condensate polaritons~\cite{PhysRevLett.109.216404, PhysRevB.91.195308, Anton_PRB2013, Gao_PRB2012, PhysRevB.85.155320} but, to our knowledge, there has been no investigation on planar microlensing. Here, we demonstrate strong focusing of polariton waves outside of their pumped regions in both the steady state, and a multi-energy-component state. The latter is characterized by tunable, and high contrast, intensity beatings at the focal point reaching frequencies as high as $250$ GHz.

The rest of the paper is organized as follows: In Sec.~\ref{Section_2}, as a preparation for the more complicated non-linear polariton system, we study microlensing in a damped two-dimensional (2D) Schr\"{o}dinger system corresponding to the linear (non-interacting) polariton regime. In Sec.~\ref{Section_3}, we nonresonantly excite a source condensate which approximately emits a coherent plane wave polariton flow. This flow impinges on a planoconcave microlens generated by a second nonresonant beam. We analyse the response of the condensate waves against this additional microlens potential and develop an argument for the operational requirements of efficient reservoir lensing. In Sec.~\ref{Section_4}, we investigate a simpler idea of using only a lens-shaped beam pumped above condensation threshold, resulting in spontaneous formation of condensate profiles strongly focused away from their pumped region. Finally, the general conclusion of our reservoir optics scheme is drawn in Sec.~\ref{Section_5}.
\section{Planar polariton microlensing in the linear regime}
\label{Section_2}
We start our analysis by considering first non-interacting (linear regime) lower branch polaritons in a planar microcavity described, in the effective mass approximation, with a 2D Schr\"{o}dinger equation with a complex potential and drive term representing a resonant laser excitation,
\begin{align} \label{2d_shcrodinger_equation}
i\hbar\frac{\partial \Psi}{\partial t}  = & \left[-\frac{\hbar^{2}\nabla^{2}}{2m}+ (V_{r} + i V_{i}) V({\mathbf{r}}) -\frac{i \hbar \gamma}{2} \right]\Psi  \nonumber
\\
&+ E(\mathbf{r})e^{-i(\omega_{s}t - \mathbf{k}_s \cdot \mathbf{r})}.
\end{align}
Here, $m$ is the effective polariton mass, $\gamma$ is the linear decay rate due to the lossy cavity mirrors, $V_r$ and $V_i$ quantify the the real and imaginary parts of the potential whose spatial profile $V(\mathbf{r})$ is taken to have a step-function boundary for brevity, and $E(\mathbf{r})$ is a coherent (resonant) driving field with frequency $\omega_{s}$ and wavevector $\mathbf{k}_s$. 

We will focus on steady state solutions  $\Psi(\mathbf{r},t) = \psi(\mathbf{r}) e^{-i \omega_s t}$ giving the time-independent Schr\"{o}dinger equation,
\begin{equation} 
\hbar\omega_s \psi =  \left[-\frac{\hbar^{2}\nabla^{2}}{2m}+ (V_r + i V_i)V({\mathbf{r}}) -\frac{i\hbar \gamma}{2} \right]\psi  + E(\mathbf{r})e^{i \mathbf{k}_s \cdot \mathbf{r}}.
\end{equation}
Inside the potential $V(\mathbf{r})$ we obtain the homogeneous Helmholtz equation,
\begin{eqnarray}
& \nabla^{2}\psi+k^2 \psi= 0,
\label{inhomogeneous_Helmholtz_equation}
\\
& k^2 =\dfrac{2m}{\hbar^{2}}\left[\hbar\omega_{s}- V_r - i\Big(V_i-\dfrac{\hbar \gamma}{2}\Big)\right].
\label{wavevector_of_the_lens}
\end{eqnarray}
Under resonant driving, $k_{s}=\sqrt{2m\omega_{s}/\hbar}$, the refractive index of the complex-valued potential with respect to the source is,
\begin{eqnarray}
n^{\prime}=\sqrt{1-\frac{V_r}{\hbar\omega_{s}}-\frac{i}{\hbar\omega_{s}}\Big(V_i-\frac{\hbar\gamma}{2}\Big)}.
\label{refractive_index}
\end{eqnarray}
Assuming $V_r<\hbar\omega_{s}$ and $V_i-\hbar\gamma/2 \ll \hbar\omega_{s}-V_r$ we can Taylor expand Eq.~\eqref{refractive_index} so it reads,
\begin{equation}
 n^{\prime}=n+i\kappa,
\end{equation}
where
\begin{equation}
n=\sqrt{1-\dfrac{V_r}{\hbar\omega_{s}}}, \qquad \kappa=-\dfrac{2V_i-\hbar\gamma}{4\hbar\omega_{s}n}.
\label{refractive_index_simplified}
\end{equation}
For a planoconcave shaped potential $V(\mathbf{r})$ whose edge is depicted with a green solid line in Figs.~\ref{fig2_b} and~\ref{fig2_e} (see Supplemental Material for the case of a positive meniscus lens) we recall the Lensmaker's equation in the ray optics limit where the focal length $f$ follows,
\begin{eqnarray}
f=\frac{R}{1-n}.
\label{flocal_length_planoconcave}
\end{eqnarray}
Here, $R$ is the radius of curvature of the back surfaces of the lens. In the case of planar microlenses, whose characteristic spatial scale is only several wavelengts, the focal length will deviate from Eq.~\eqref{flocal_length_planoconcave} due to pronounced scattering and interference of the waves impinging on the lens. We thus numerically solve the steady of Eq.~\eqref{2d_shcrodinger_equation} under resonant excitation of plane waves that pass through the planoconcave microlens. We base our parameters on state of the art inorganic microcavities for generating polariton condensates~\cite{cilibrizzi2014polariton}: $\gamma^{-1} = 5.5 \mathrm{\,ps}$ and $m=4.9 \times 10^{-5} m_0$ where $m_0$ is the free electron mass.
\begin{figure*}[t]
\includegraphics[scale=1.0]{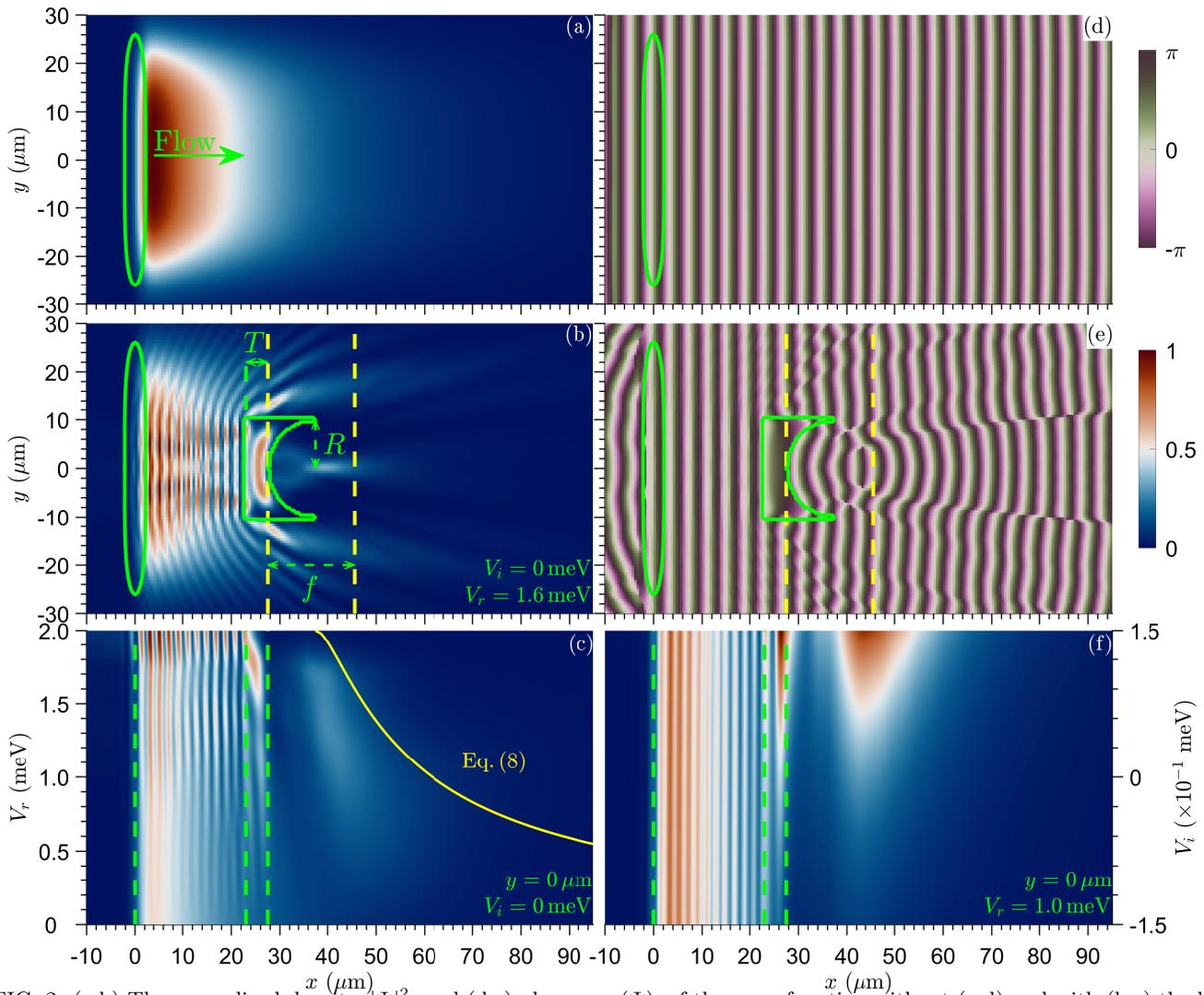}
\phantomsubfloat{\label{fig2_a}}
\phantomsubfloat{\label{fig2_b}}
\phantomsubfloat{\label{fig2_c}}
\phantomsubfloat{\label{fig2_d}}
\phantomsubfloat{\label{fig2_e}}
\phantomsubfloat{\label{fig2_f}}
\vspace{-4\baselineskip}
\caption{(a,b) The normalized density, $|\Psi|^2$, and (d,e) phase, $\text{arg}{(\Psi)}$, of the wave function without (a,d) and with (b,e) the lens potential. The source and the lens are outlined with green solid lines. (c,f) Line profile of the wavefunction density along the lens axis ($y=0$) as a function of varying real (c) and imaginary (f) part of the potential. In (b) and (e) the yellow dashed lines indicate the principal plane of the lens and the focal point with distance $f$. The green vertical dashed lines in (c) and (f) indicate the source location and the front and back surface of the lens.}
\label{fig2}
\end{figure*}

Since $n=1$ and $n<1$ outside and inside the lens, respectively, an incident planar wavefront from the left will transmit through the lens to converge into a cylindrical wavefront on the right side. We numerically solve for the steady states of Eq.~\eqref{2d_shcrodinger_equation} under resonant excitation at the left side of the lens with a profile, $E(\mathbf{r})=\mathrm{exp}[-x^2/(2\sigma^{2}_{x})-y^2/(2\sigma^{2}_{y})]$ whose full-width-half-maximum (FWHM) is outlined with green in Fig.~\ref{fig2_a} centered at $x=0$. We set the energy of the source excitation to $\hbar\omega_{s}=\SI{2.0}{\mathrm{meV}}$ so to have a rapidly varying phase front and remaining within the parabolic (dispersion) regime. The corresponding steady state density and phase profiles of $\Psi(\mathbf{r})$ without any lens potential are shown in Figs.~\ref{fig2_a} and~\ref{fig2_d}.

When a planoconcave potential $V(\mathbf{r})$, of size in the order of several wavelengths ($\lambda_s \approx 3.9$ $\mu$m), is introduced both transmitted and scattered waves contribute in a complicated way to the focal region on the right side of the lens [see Fig.~\ref{fig2_b}]. Here we  set $R= \SI{10.0}{\mu m}$ and the lens' thickness $T= \SI{4.5}{\mu m}$ which corresponds to $n\approx0.4472$ and $f\approx\SI{18.1}{\mu m}$ indicated by the yellow dashed lines in Figs.~\ref{fig2_b} and~\ref{fig2_e}. We observe a focal region (i.e., the white-ish region of converged/focused waves) that lies outside the lens curvature $R$ and within the ray-optics focal length $f$ [Eq.~\eqref{flocal_length_planoconcave}], as a consequence of the microscopic nature of the lens shape. We stress that the low polariton intensity in the focal region is dominated by their rapid decay rate $\gamma$ used in our simulation. However, condensation of polaritons with large lifetimes reaching $\gamma^{-1} = 270$ ps has also been demonstrated~\cite{Sun_PRL2017}, leading to longer propagation lengths and timescales to manipulate the condensate flow.

In Fig.~\ref{fig2_c}, we show the line profile of the wavefunction density at $y=0$ for varying real potential strength $V_r$. The focal region both shrinks and the focal length decreases as the potential strength increases in qualitatively agreement with Eq.~\eqref{flocal_length_planoconcave} (yellow solid line). It is worth mentioning that from Eq.~\eqref{refractive_index_simplified} one can, in principle, achieve epsilon-near-zero ($n=0$) lensing, which has been studied extensively in metamaterials~\cite{Alu_PRB2007}, by tuning the excitation frequency. However, at $\hbar \omega_s \approx V_r$ incident waves undergo stronger reflection leading to a pronounced interference pattern like seen in Fig.~\ref{fig2_b} to the left of the lens. We also vary the imaginary part of the potential $V_{i}$ in Fig.~\ref{fig2_f} showing a clear amplification of the transmitted waves in accordance with the imaginary part of the refractive index in Eq.~\eqref{refractive_index_simplified}.

\section{Planar reservoir microlensing with polariton condensates}
\label{Section_3}
\subsection{Generalized Gross-Pitaevskii model}
\label{sec.GP}

Having characterized the effects of the 2D planoconcave microlens on an incoming plane wave, we now move to the nonlinear regime with condensates of polaritons. The polariton condensate wavefunction $\Psi(\mathbf{r},t)$ obeys a generalized Gross-Pitaevskii equation coupled to a driven exciton reservoir $N(\mathbf{r},t)$ rate equation~\cite{PhysRevLett.99.140402},
\begin{eqnarray}
i\hbar\frac{\partial \Psi}{\partial t}&=&\bigg[-\frac{\hbar^{2}\nabla^{2}}{2m}
 +\alpha |\Psi|^2 +G\left(N + \frac{\eta P(\mathbf{r})}{\Gamma}\right) \nonumber 
\\
&&  +\frac{i \hbar}{2}\left(\xi N-\gamma\right)\bigg]\Psi,
\label{gross-pitaevskii_equation}
\\
\frac{\partial N}{\partial t}&=&-\left(\Gamma+\xi |\Psi|^{2}\right)N +P(\mathbf{r}).
\label{rate_equation_of_reservoir}
\end{eqnarray}
Here, $G = 2 g |\chi|^2$ and $\alpha = g |\chi|^4$ are the polariton-reservoir and polariton-polariton interaction strengths, respectively, $g$ is the exciton-exciton dipole interaction strength, $|\chi|^2$ is the excitonic Hopfield fraction of the polariton, $\xi$ is the scattering rate of reservoir excitons into the condensate, $\Gamma$ is the reservoir decay rate, $\eta$ quantifies additional blueshift coming from a dark background of excitons which do not scatter into the condensate, and $P(\mathbf{r})$ is the nonresonant continuous-wave pump. The parameters used in all simulations are based on negatively detuned cavities, $|\chi|^{2}=0.4$, with GaAs-type quantum wells, $g= 1 \, \mu\mathrm{eV\,\mu m^{2}}$. Remaining parameters are taken similar to those used to describe past experiments, 
$\hbar \xi=2.8g$; 
$\eta=5$; and $\Gamma = \gamma$.

Let us quantify the nonresonant pump as $P(\mathbf{r}) = P_0 f(\mathbf{r})$ where $P_0$ is a positive scalar denoting the power density of the pump laser and $f(\mathbf{r})$ is its profile. It is instructive to define the condensation threshold which, formally, is a bifurcation point separating the so-called normal (uncondensed) state ($|\Psi|=0$) and the condensed state ($|\Psi| \neq 0$). The threshold can be identified as the point where a single frequency component of our system in the linear regime crosses from negative to positive imaginary value (i.e., small $|\Psi|$ starts growing exponentially in time). Alternatively, one can also estimate the threshold of Eq.~\eqref{gross-pitaevskii_equation} numerically by expanding the reservoir steady state,
\begin{equation}
    N = \frac{P(\mathbf{r})}{\Gamma + \xi |\Psi|^2} = \frac{P(\mathbf{r})}{\Gamma} \left[1  - \frac{\xi |\Psi|^2}{\Gamma} + \mathcal{O}{(|\Psi|^4)} \right]
\end{equation}
and compare the contribution between the zeroth and the first order terms. Integrating through space we can write the following inequality:
\begin{equation}
\frac{\xi}{\Gamma} \int f(\mathbf{r}) |\Psi|^2 \, d\mathbf{r} < \epsilon.
\end{equation}
Here, $\epsilon \ll 1$ is some small, reasonably chosen, numerical tolerance to determine the threshold. Physically, the above expression simply states that around threshold any nonlinear effects on the reservoir are small. In this weak nonlinear regime the potential generated by the pump is approximately,
\begin{equation} \label{eq.pot}
V(\mathbf{r}) \simeq \frac{P(\mathbf{r})}{\Gamma} \left[ G\left( 1 + \eta \right) + i\frac{ \hbar  }{2 }\xi\right]. 
\end{equation}
Separating the real and imaginary parts gives,
\begin{equation} \label{eq.Vr_Vi}
 V_r = \frac{P_0}{\Gamma}G (1 + \eta), \qquad  V_i = \frac{P_0 \hbar \xi}{2 \Gamma}.
\end{equation}
For a homogeneous pump $P(\mathbf{r}) = P_0$ the threshold power corresponds to the balance of gain and dissipation $V_i - \hbar \gamma/2 = 0$ which gives $P_{0,th} = \gamma \Gamma/ \xi$. For inhomogeneous pump spots the threshold power is bigger due to additional planar losses of waves from the spatially finite gain region.

The pump which induces the lens potential is constructed using Gaussian blurring on a step function which naturally mimics the limited resolution of using spatial light modulators in experiment and finite exciton diffusion,
\begin{equation}
    f(\mathbf{r}) = \int F(\mathbf{r}') e^{-|\mathbf{r} - \mathbf{r}'|^2/2w^2} \, d\mathbf{r}' 
\label{Gaussian_bluring}
\end{equation}
where
\begin{equation}
    F(\mathbf{r}) = 
    \begin{cases}
  1 & \text{for } \mathbf{r} \in \mathcal{L}\\    
  0 & \text{else}.    
\end{cases}
\end{equation}
where $\mathcal{L}$ is the lens area. For all lens shapes used in calculations of Figs.~\ref{fig4}-\ref{fig5_v2} we apply a Gaussian blur corresponding to $w \approx \SI{0.85}{\mu m}$ ($\SI{2.0}{\mu m}$ FWHM).

\subsection{Numerical results on reservoir lensing}
We will consider two separate pumps $P(\mathbf{r}) = P_S(\mathbf{r}) + P_L(\mathbf{r})$, of characteristic sizes $D_S$ and $D_L$, which are referred to as the {\it source} and the {\it lens} as introduced in Sec.~\ref{Section_2} and depicted in Fig.~\ref{fig1}. We will denote the complex-valued potential coefficients for the source and the lens potentials as $V_S = V_{r,S} + i V_{i,S}$ and $V_L= V_{r,L} + i V_{i,L}$, respectively. Conservation of energy tells us that polaritons generated at the source will obtain kinetic energy following: 
\begin{equation}
V_{r,S} =\frac{2 \hbar^2 \pi^2 }{ m \lambda^2}.
\end{equation}
Let us list some requirements in order to obtain steady state lensing of polariton waves with wavelength $\lambda$ coming from the source and passing through the lens:
\begin{figure*}[t]
\centering
\includegraphics[scale=1.0]{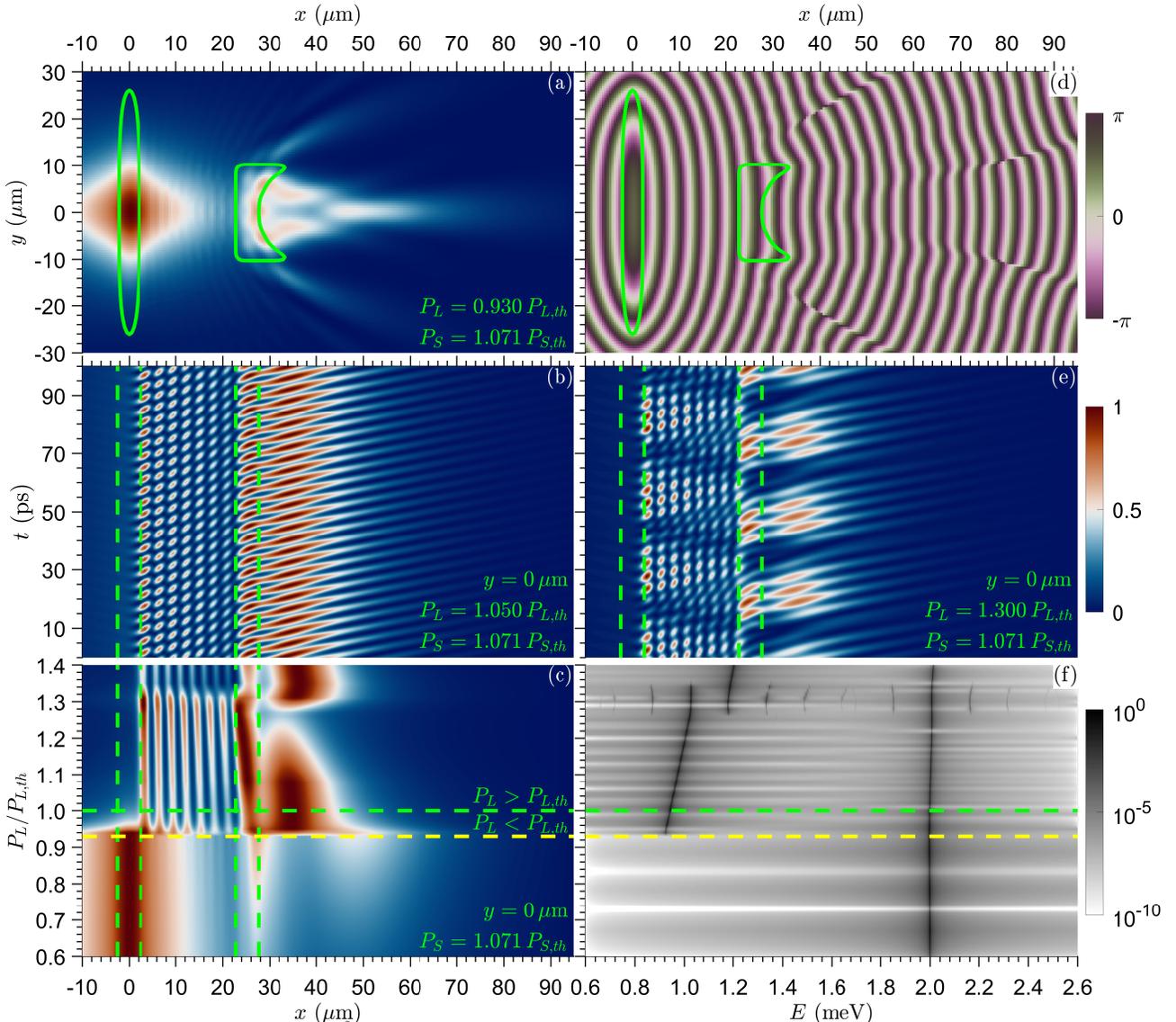}
\phantomsubfloat{\label{fig4_a}}
\phantomsubfloat{\label{fig4_b}}
\phantomsubfloat{\label{fig4_c}}
\phantomsubfloat{\label{fig4_d}}
\phantomsubfloat{\label{fig4_e}}
\phantomsubfloat{\label{fig4_f}}
\vspace{-4\baselineskip}
\caption{(a) Normalized density $|\Psi|^2$ and (d) phase map $\text{arg}{(\Psi)}$ of the condensate in the steady state under nonresonant pumping. (b,e) Time-resolved density line profile along $y=0$ for two different pump powers. (c) Time-integrated density line profile, and (f) corresponding spectral weight for varying lens power. Data is normalized at each step in $P_{L}/P_{L,th}$. The FWHM of the pump profiles is outlined with green solid lines. Vertical green dashed lines indicate the outer and inner boundary of the source and lens at $y=0$. The horizontal green dashed line indicates the threshold of the isolated lens. The horizontal yellow dashed line indicates the onset of periodic dynamics characterized by more than one spectral peak.}
\label{fig4}
\end{figure*}

\begin{enumerate}
\item[(i)] $D_L > \lambda$, the lens has to be large enough to refract the incident waves.
\item[(ii)] $0< V_{r,S} - V_{r,L} = \Delta$, waves must be propagating in the lens.
\item[(iii)] $P_L < P_{L,\text{th}}$, the lens should be below threshold.
\item[(iv)] $P_S > P_{S,\text{th}}$, the source must be above threshold.
\end{enumerate}
Here, $P_{S(L),\text{th}}$ are the threshold powers of the isolated source (lens) pumps. 

We can rewrite requirements (i) and (ii) in terms of the model parameters, respectively,
\begin{align}
& D_L > \hbar \pi \sqrt{\frac{2 \Gamma}{m P_{0,S} G (1 + \eta)}} = \lambda,
\\
& 0 < \frac{P_{0,S} - P_{0,L}}{\Gamma} G (1 + \eta) = \Delta. \label{eq.Delta}
\end{align}
Here, $P_{0,S(L)}$ denotes the power density of the nonresonant source (lens) pump. It therefore becomes evident that increasing $P_{0,S}$ will allow us to satisfy both requirements. However, $\Delta$ needs to be reasonably bounded to obtain good focusing of transmitted waves. This is evident from the variable maximum intensity in the focal region in Fig.~\ref{fig2_c}. Therefore, arbitrarily increasing $P_{0,S}$ does not guarantee good focusing of polariton waves. We also note that requirement (iii) is not strict as we will see later.

We demonstrate our reservoir lensing scheme in Fig.~\ref{fig4} by numerically solving the generalized Gross-Pitaevskii and reservoir model. We set the profile of the source pump to be cigar-shaped to approximately generate plane waves $f_{S}(\mathbf{r})=\mathrm{exp}[-x^2/(2\sigma^{2}_{x})-y^2/(2\sigma^{2}_{y})]$ in which $\sigma_{x} \ll \sigma_y$. The lens is taken to be planoconcave shaped with $R=\SI{10.0}{\mu m}$ and $T = \SI{4.5}{\mu m}$. The FWHM of the source and lens are outlined with green solid curves in Figs.~\ref{fig4_a} and~\ref{fig4_d}. We stress that due to the different profiles of the source and the lens their threshold powers are different. 

One of the main differences between the resonant scheme discussed in Sec.~\ref{Section_2} and the current nonresonant scheme is the vivid localization of the source condensate along the vertical direction shown in Fig.~\ref{fig4_a}. This effect stems from the anisotropic gain region favoring modes with minimal losses, and effective attractive interactions between the condensate and the reservoir due to the gain-saturation mechanism ~\cite{Smirnov_PRB2014}. None-the-less, enhancement of propagating waves in the focal region can be observed clearly in Fig.~\ref{fig4_a}, partly due to amplification from the lens gain. The phase map shown in Fig.~\ref{fig4_d} is very different from that in Fig.~\ref{fig2_d} which stems from the large detuning between the source waves and the lens potential in simulation, i.e. $\Delta = V_{r,S} - V_{r,L} \approx 2.0 \text{ meV} - 0.8 \text{ meV} = 1.2\text{ meV}$. In order to reduce the detuning $\Delta$, and get stronger focusing, one could pump the lens harder. However, this triggers condensation inside the lens and requirement (iii) is violated. Moreover, reinforcing ``ballistic" interactions between the source and the lens region have lowered the lens threshold~\cite{Aleiner_PRB2012, Khan_PRA2016} (yellow dashed line in Fig.~\ref{fig4}). These complex wave dynamics make it therefore a nontrivial task to adjust the detuning arbitrarily $\Delta$ to obtain stronger focusing while---at the same time---keeping the lens pump below threshold.

If, on the other hand, requirement (iii) is relaxed and the lens power is made variable then interesting nonlinear physics become enhanced. In Fig.~\ref{fig4_c} we show the time-integrated line profile of the wavefunction density at $y=0$ for varying lens power $P_L$, and the corresponding energy spectrum in Fig.~\ref{fig4_f}. As discussed at the start of the section, the system favors a steady state behaviour when $P_L$ is small, characterized by a single clear spectral line in Fig.~\ref{fig4_f}. In this regime, the results are similar to those of a static lens potential impinged by resonantly excited waves discussed in Sec.~\ref{Section_2}. However, as the lens power increases, an additional spectral line appears and nonstationary periodic solutions form as a results of intricate interactions between the condensate polaritons generated at the source and the lens, in agreement with recent experiments~\cite{topfer2020time}. An example of two such solutions in the time domain is shown in Figs.~\ref{fig4_b} and~\ref{fig4_e}. Clear $\approx 252$ GHz intensity beatings in the focal region can be observed in Fig.~\ref{fig4_b} whereas Fig.~\ref{fig4_e} shows two dominant beat frequencies.

\section{Reservoir lenses above threshold}
\label{Section_4}
\begin{figure}[t!]
\includegraphics[scale=1.0]{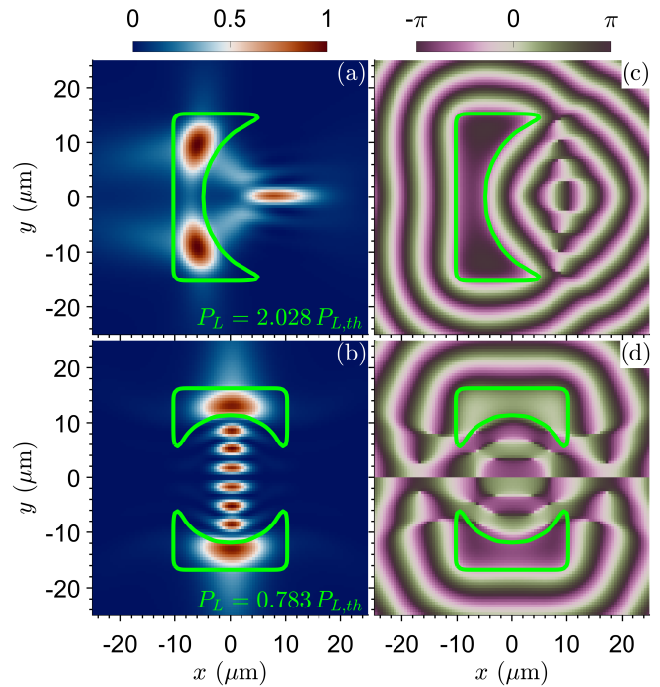}
\phantomsubfloat{\label{fig5_a}}
\phantomsubfloat{\label{fig5_b}}
\phantomsubfloat{\label{fig5_c}}
\phantomsubfloat{\label{fig5_d}}
\vspace{-2.5\baselineskip}
\caption{(a,b) Steady state condensate density $|\Psi|^2$ and (c,d) phase $\text{arg}{(\Psi)}$ for two different nonresonant pump configurations. The nonresonant pump is shaped into a (a) planoconcave lens showing clear focusing of the emitted waves outside the pumping area, and a (b) planoconcave resonator made from two lenses (emitter) facing each other. Note that each individual lens is below threshold but the system/resonator as a whole has a lower threshold and thus supports a standing wave condensate at lower powers. The FWHM of the pump profiles is outlined in green.}
\label{fig5}
\end{figure}

There are limitations to the source and lens scheme in previous section which cannot be quantified nicely given the complex wave dynamics at play. Firstly, reinforcing behaviour between the source and the lens regions results in lowered threshold gain of the interacting system which can lead to condensation into extended quasinormal standing wave modes that are supported by both the source and the lens region. This is a general feature of interacting dissipative systems, such as coupled lasers, or interacting polariton condensates~\cite{Aleiner_PRB2012, Khan_PRA2016}.
Second, the source pump size would, in general, need to be larger than the lens in order to avoid $\Delta$ getting too large (i.e., smaller source pumps need to be driven with higher power and thus emit waves with higher energy). This can lead to thermally induced self-trapping of the source condensate~\cite{Ballarini_PRL2019}.

To overcome these issues, we consider a more simple case where the source pump $P_S(\mathbf{r})=0$ is omitted and just the lens $P_L(\mathbf{r})$ is driven above threshold. Indeed, the lens region then plays the role of a carefully designed anisotropic planar emitter from which waves radiate to constructively interfere. In Fig.~\ref{fig5} we show the condensate steady state for a pump profile shaped into a planoconcave lens and driven above threshold. Polariton waves generated in the pump region are propagating along the direction normal to the lens surface and form a strong focal region with a clear phase shift.

\begin{figure}[t!]
\includegraphics[scale=1.0]{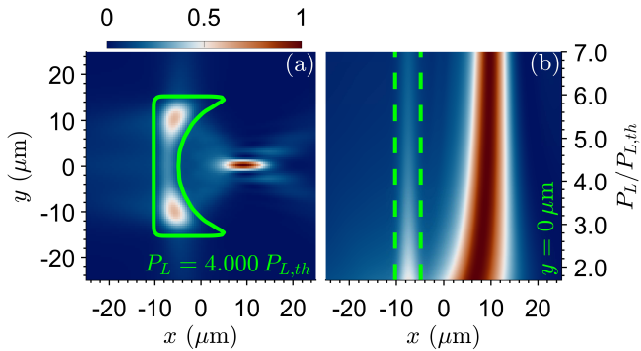}
\phantomsubfloat{\label{fig_line_profile_a}}
\phantomsubfloat{\label{fig_line_profile_b}}
\vspace{-2.5\baselineskip}
\caption{(a) Condensate density for a planoconcave shaped nonresonant pump at high powers and (b) line profile along $y=0$ for varying power. The vertical green dashed lines indicate the pumped region.}
\label{fig5_v2}
\end{figure}

When the ``lens" power is increased then the contrast between the condensate density within and outside the lens region increases as shown in Figs.~\ref{fig_line_profile_a} and~\ref{fig_line_profile_b}, where in the latter we plot the condensate density line profile along the lens axis~\cite{footnote1}. These results underpin the potential of using anisotropic shaped nonresonant excitation beams to generate high density polariton condensates spatially separated from any influence of the background exciton reservoir such as strong dephasing or spatial hole burning effects. In the Supplemental Material we also provide results on a pump shaped into a positive meniscus lens.

We also investigate the potential of designing planar resonators by setting two identical lens-shaped pump profiles facing each other [see Figs.~\ref{fig5_b} and~\ref{fig5_d}]. Now, a clear condensate standing wave forms, strongly localized along the horizontal direction. The mode number of this planar standing wave can easily be tuned by changing the distance between two lenses or their pump power~\cite{topfer2020time}. Moreover, the pump polarization of each pump lens can also be adjusted to design condensate standing waves with intricate polarization patterns~\cite{Pickup_PRB2021}. These results open a pathway in generating structured, high density, polariton condensates spatially separated from the direct influence of the reservoir by simply adjusting the geometric configuration and the excitation power of the nonresonant pump.

\section{Conclusions}
\label{Section_5}

In summary, we have theoretically investigated all-optical planar microlensing techniques on condensates of exciton-polaritons. The lenses are created by using spatially patterned nonresonant excitation profiles that provide both gain and blueshift to the polariton modes. We stress that our scheme should not be confused with that of resonant control~\cite{Amo_PRB2010, Sanvitto_NatPho2011} where auxiliary ``condensates" are directly injected to provide spatially patterned polariton blueshift.

We studied the condensate dynamics first in a source-and-lens pump setup as shown in Fig.~\ref{fig1}. We provide a comparison of the rich nonlinear dynamics to that of linear Schr\"{o}dinger wave mechanics. Scanning across pump power parameters, we demonstrate a departure of the condensate steady state, resembling the linear Schr\"{o}dinger dynamics, into a stable limit cycle state characterized by multiple spectral peaks and rapid ($\approx 252$ GHz) density oscillations in the focal region. This result holds promises for polaritonic clock generators in integrated circuits~\cite{Leblanc_PRB2020}. We next studied the condensate behaviour in a simpler setup consisting of only a single lens shaped pump driven above threshold. This resulted in highly anisotropic condensate emission and, amazingly, much stronger focusing of condensate waves outside the pump region, as compared to the source-and-lens scheme. This last result opens possibilites in generating polariton condensates that are separated from detrimental reservoir dephasing effects and might obtain unprecedented coherence times, favorable for highly sensitive planar matter-wave interferometers~\cite{Sturm_NatComm2014}.

The possible reservoir devices and their applications are not limited by the examples we present in this paper, and we hope this work will stimulate the theoretical and experimental application of reservoir optics in polariton condensates. Our findings are also relevant to atomtronics~\cite{Li_NatComm2019, Pandey_PRL2021} where arbitrary all-optical control over the atom's potential landscape is possible~\cite{Nogrette_PRX2014}. 

\begin{acknowledgments}
The authors acknowledge the support of the UK’s Engineering and Physical Sciences Research Council (grant EP/M025330/1 on Hybrid Polaritonics), and the European Union’s Horizon 2020 program, through a FET Open research and innovation action under the grant agreement No. 899141 (PoLLoC). H.S. acknowledges the Icelandic Research Fund (Rannis), grant No. 217631-051. Y.W.’s studentship was financed by the Royal Society RGF$\backslash$EA$\backslash$180062 grant.
\end{acknowledgments}

\bibliography{references}

\widetext
\clearpage
\begin{center}
\textbf{\large Reservoir optics with exciton-polariton condensates: Supplemental information}
\end{center}
\setcounter{equation}{0}
\setcounter{figure}{0}
\setcounter{table}{0}
\setcounter{page}{1}
\makeatletter
\setcounter{section}{0}
\renewcommand{\thesection}{S\arabic{section}}
\renewcommand{\theequation}{S\arabic{equation}}
\renewcommand{\thefigure}{S\arabic{figure}}
\renewcommand{\bibnumfmt}[1]{[S#1]}
\renewcommand{\citenumfont}[1]{S#1}

\section{Positive meniscus lens in the linear regime}

\begin{figure*}[t]
\includegraphics[scale=1.0]{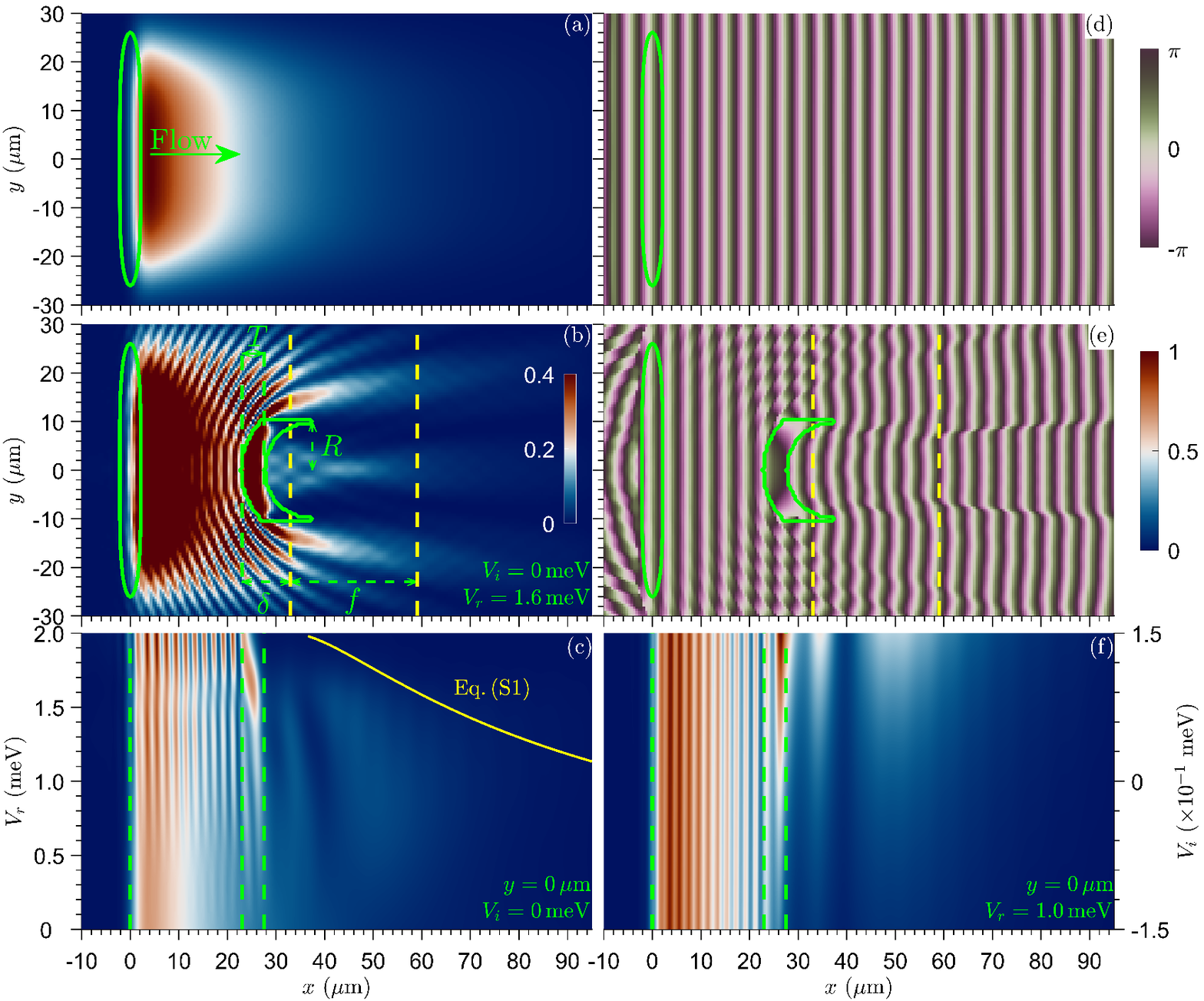}
\phantomsubfloat{\label{Figure_S1a}}
\phantomsubfloat{\label{Figure_S1b}}
\phantomsubfloat{\label{Figure_S1c}}
\phantomsubfloat{\label{Figure_S1d}}
\phantomsubfloat{\label{Figure_S1e}}
\phantomsubfloat{\label{Figure_S1f}}
\caption{(a,b) The normalized density, $|\Psi|^2$, and (d,e) phase, $\text{arg}{(\Psi)}$, of the wave function without (a,d) and with (b,e) the lens potential. The source and the lens are outlined with green solid lines. (c,f) Line profile of the wavefunction density along the lens axis [$y=0$] as a function of varying real (c) and imaginary (f) part of the potential. In (b) to detail the focal region, the value of the normalized density larger than $0.4$ is saturated in the colormap. In (b) and (e) the yellow dashed lines indicate the principal plane of the lens and the focal point with distance $f$ and $\delta$ is the distance between the front surface to the principal plane (the first yellow dashed line from left to right). The green vertical dashed lines in (c) and (f) indicate the source location and the front and back surface of the lens and $T$ indicates the distance between two green dashed lines.
}
\label{Figure_S1}
\end{figure*}

For a double concave shaped potential $V(\mathbf{r})$, we recall the Lensmaker's equation in the ray optics limit where the focal length $f$ follows,
\begin{equation}
\frac{1}{f}=(n-1)\left[ \frac{1}{R_{1}}-\frac{1}{R_{2}}+\frac{(n-1)T}{nR_{1}R_{2}} \right].
\label{lensmakers_equation}
\end{equation}
Here, $R_{1,2}$ are the radius of curvature of the front and the back surfaces of the lens, respectively (left and right edges), and $T$ represents the lens' thickness. In the case of using a positive meniscus lens where $R_1>0$, the thickness $T$ of the lens in Eq.~\eqref{lensmakers_equation} needs to be taken into account. It is instructive to define the distance,
\begin{equation}
\delta=f\frac{T}{R_{1}}\left(\frac{1}{n}-1\right),
\label{principal_plane}
\end{equation}
which refers to the distance between the front surface in the positive meniscus lens and the principal point [see Fig.~\ref{Figure_S1b}]. The radius of the front and back surface of the lens is, respectively, $R_{1}=\SI{14.5}{\mu m}$ and $R_{2}=\SI{10.0}{\mu m}$, and the thickness of the lens $T=\SI{4.5}{\mu m}$. The radius of the back surface, the lens' thickness, and the resonant source is the same one used in Fig.~2 in main text. For a lens with $V_{r}=\SI{1.6}{\mathrm{meV}}$ and $V_{i}=\SI{0}{\mathrm{meV}}$, we obtain $\delta\approx\SI{10.0}{\mu m}$ and $f\approx\SI{26.1}{\mu m}$ using Eq.~(7) in the main text to get the effective refractive index $n$.

Figure~\ref{Figure_S1} shows the same numerical experiment as Fig.~2 in the main text using this time a positive meniscus lens. Compared to the results of planoconcave lens (main text) we observe enhanced scattering of the incident plane wave front onto to the positive meniscus which results in poorer focused transmission. This result is in contrast to the case presented in Fig.~\ref{Figure_S3} where a positive meniscus emitter focuses waves more efficiently than a planoconcave lens.
\section{Reservoir positive meniscus lenses above threshold}
\begin{figure}[t!]
\includegraphics[scale=1.0]{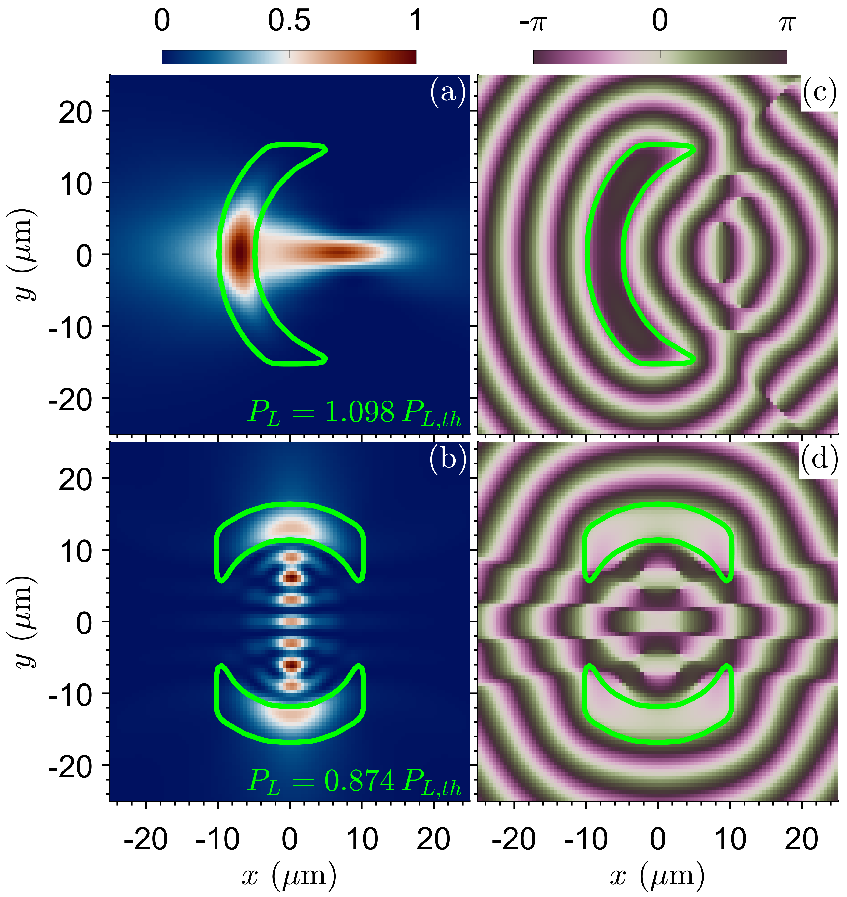}
\phantomsubfloat{\label{Figure_S2a}}
\phantomsubfloat{\label{Figure_S2b}}
\phantomsubfloat{\label{Figure_S2c}}
\phantomsubfloat{\label{Figure_S2d}}
\caption{(Left column) Condensate density $|\Psi|^2$ and (Right column) phase $\text{arg}{(\Psi)}$ in the steady state. The nonresonant pump is shaped into a (a) positive meniscus lens showing clear focusing of the emitted planar waves outside the pumping area, and a (b) positive meniscus resonator made from two lenses facing each other. Note that each individual lens is below threshold but the system/resonator as a whole has a lower threshold and thus supports a standing wave condensate at lower powers. The FWHM of the pump profiles is outlined in green.}
\label{Figure_S2}
\end{figure}

\begin{figure}[t!]
\includegraphics[scale=1.0]{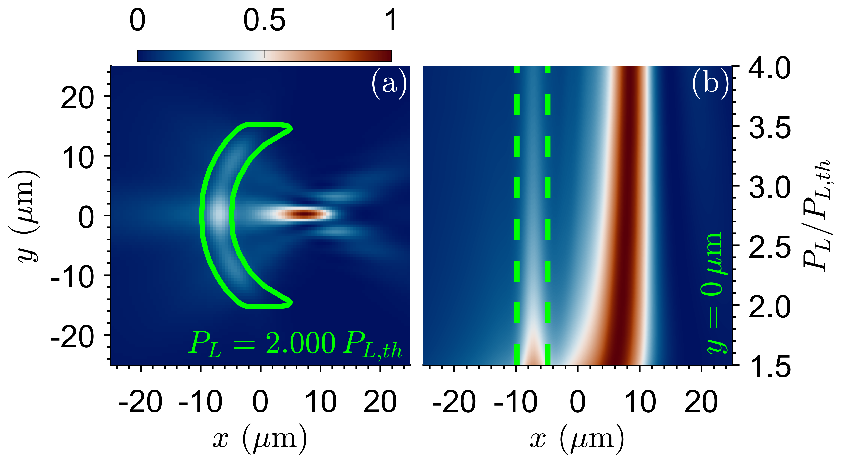}
\phantomsubfloat{\label{Figure_S3a}}
\phantomsubfloat{\label{Figure_S3b}}
\caption{(a) Same as Figs.~\ref{Figure_S2a} for higher lens powers and $\eta = 180$ and $\hbar \xi = 84.2g$. (b) Condensate density line profile along $y=0$ for varying lens power corresponding to panels (a). The vertical green dashed lines indicate the pump (lens) region.}
\label{Figure_S3}
\end{figure}

Figures~\ref{Figure_S2a} and~\ref{Figure_S3} show the same calculation as given in Fig.~4 and~5 (planoconcave lens) in the main text with this time a pump shaped into a positive meniscus lens. In comparison with the planoconcave lens detailed in main text, we point out the stronger localization and higher density of the condensate outside the pump region in the positive meniscus lens.

\end{document}